\documentclass   [a4paper,12pt]{article}
\usepackage{graphicx}
\begin{document}
   
\begin{center}
{\bf\Large Elastic pp scattering and the internal structure of colliding protons }   \\[2mm]
    Milo\v{s} V. Lokaj\'{\i}\v{c}ek, Vojt\v{e}ch Kundr\'{a}t

Institute of Physics, v.v.i., Academy of Sciences of the Czech Republic,\\
 18221 Prague, Czech Republic
\end{center}

Abstract

Elastic scattering data gained for proton-proton collisions at high energies are being analyzed at the present practically only with the help of purely phenomenological mathematical models. And the question exists whether in the impact parameter plane the elastic processes may be interpreted as central or peripheral. From realistic point of view the peripherality should be preferred and one may expect that nucleon dimensions should manifest in some way in elastic data.
It will be demonstrated that the elastic pp data at the energy of $53\; GeV$  in the interval $|t|\in (0., 4.)\; [GeV^2]$ may be interpreted well as the superposition of mutual collisions of two internal structures with maximal external dimensions; the approximate dimensions and frequencies of corresponding structures being established on the basis of measured elastic data. 
 \\ 

\section {Introduction}

Elastic differential cross sections of two colliding protons at high energies represent probably the most precisely established nucleon data. Only different phenomenological models of elastic processes are, however, available, even if a series of models interpreting nucleons as consisting of quarks and partons have been applied to in the case of non-elastic collisions. 
Similar models have not been successful in describing elastic collisions. 

It follows from the analysis based on eikonal model that the peripheral behavior of elastic processes should be acceptable and even partially preferred \cite{kun}. In such a case one should expect that mainly external nucleon dimensions might influence significantly these processes. And we may ask whether the external dimensions of different internal structures may manifest in measured elastic nucleon data.

To answer at least partially the given question we will attempt to describe collision process at the difference to the standard quantum-mechanical model  on purely probabilistic basis only. We shall assume that the 
nucleon structure is changeable and that a nucleon may exist in different internal states that may have divers dimensions.
In the peripheral alternative mainly the states exhibiting maximal external dimensions should manifest in elastic collisions. And the resulting collision characteristics should be interpreted as a sum of those belonging to individual collision types (between different pairs of nucleon states). And an interesting picture may be gained if one attempts to interpret experimental elastic data on such a simple probabilistic basis.  

In the following we shall analyze the elastic data obtained in the case of pp collisions at energy of $53\, GeV$ (ISR); see, e.g., their earlier phenomenological eikonal model analysis in Ref. \cite{kun}. In the following we shall limit ourselves, however, as already mentioned to the purely probabilistic description of different collision types in the dependence on impact parameter values $b$.
And it will be shown that the basic part of elastic data corresponding to the lower values of momentum transfer, i.e. the data in the interval  
$|t|\in (0.,4.)\, [GeV^2]$, may be well interpreted as the superposition of mutual collisions of two different internal proton states (or structures) exhibiting maximal dimensions. 
The given results may be helpful in modeling the given collision processes on  deeper grounds.
 
\section {Different internal structures of colliding protons  }

There is not any doubt that protons should be regarded as complex objects that consist of smaller constituents (quarks or partons). And one must admit that their internal structures may be changeable, which may play also a role in their interactions with other microscopic objects. However, the contemporary quark model takes these protons as having an average structure that is responsible for their behavior.

 In the following we shall not consider the types and properties of corresponding basic constituents. We shall try to deal rather with some external proton characteristics that might be responsible for their elastic collisions.
We shall assume that protons are objects with changeable dimensions, being represented in principle by a kind of different oscillating ellipsoids. It means that collision characteristics may be represented as a collision superposition of differently behaving objects.

The number of different structures may be rather great. However, in the following we shall consider for simplicity only two different structures with maximal external dimensions that are responsible for collision data corresponding to the lowest momentum transfers (or to the lowest $|t|$ values). Let us assume that these structures will be present in any nucleon with the probabilities $p_1$ and $p_2$. It means that the corresponding measured data will be represented by a superposion of three different collision types with the probabilities 
      \[  r_1=p_1^2, \;\;r_2=2p_1p_2, \;\;r_3=p_2^2.  \]

 Let us assume further that the structures are characterized by maximal dimensions $b_1$ and $b_2$, which means that three maximal values of effective impact parameters will exist: 
        \[ B_1=b_1, \;\;B_2=(b_1+b_2)/2, \;\;B_3=b_2.  \]
The individual collision kinds will be then characterized by three functions $f_1(b), f_2(b), f_3(b)$ that represent mean values of $t$ for corresponding values of impact parameter $b$; all three functions being monotony decreasing when $b$ rises from $0$ to $B_j$. Denote then by $\bar{f}_j(t)$  the inverse functions that attribute mean values of $b$ to individual values $t$.

 It will then hold for the differential cross section (established in milibarns while the dimensions of protons are given in $fm$)
\begin{equation}
   \frac{d\sigma_{el}(t)}{dt}\,=\, \sum_1^3r_j\,F_j(t)\,.\, 10     \label{difs}
\end{equation}
where
\begin{equation}
   F_j(t)\,=\,2\pi\,g_j\Bigl( \bar{f}_j(t)\Bigr) \,\frac{d\bar{f}_j(t)}{dt}
\end{equation}
and $g_j(b)$ corresponds to the probabilistic distribution of individual impact parameter values for corresponding elastic collisions if cylindrical symmetry has been assumed.

The behavior of such two colliding protons will depend on several free parameters: probabilities $p_1$ and $p_2$, two dimension values $b_1$ and $b_2$, and on two function triples $g_j(b)$ and $f_j(b)$ (or $\bar{f}_j(t)$).  Their values or shapes should be derived from corresponding experimental data.   

\section {Mathematical model and experimental data}

At difference to monotone functions $f_j(b)$ or $\bar{f}_j(t)$ the functions $g_j(b)$ should exhibit some maximums at different places in the intervals $(0,B_j)$. These functions may be, however, interpreted always as the products of two functions that should be oppositely monotone; it is possible to write
\begin{equation}
        g_j(b)\;=\; a_j(b).e_j(b)
\end{equation}         
where $a_j(b)$ are the probabilities of any mutual particle interaction (elastic as well as inelastic) while $e_j(b)$ represents the fraction of elastic processes from all possible interactions at a given impact parameter.
 
 Consequently, we have now three free monotone functions, for which the following parameterization have been chosen: 
\begin{eqnarray}                 
     a_j(b) \;&=&\; \;\;\;\;\;\;1  \hspace{35mm} (b \leq d_j),         \nonumber     \\                \;&=&\;(\cos[\frac{\pi}{2}(\frac{b-d_j}{B_j-d_j})
               ^{\delta_j}])^{\varepsilon_j}  \hspace{10mm} (b > d_j), \label{ajb} \\
     e_j(b)\;&=&\; e^{-(h_j(B_j-b))^{\theta_j}}\frac{1+c_j}
                              {1+c_je^{-(h_j(B_j-b))^{\theta_j}}}  \,.  \label{ejb}     \end{eqnarray}                 
And similar parametrization will be made use of for functions $f_j(b)$ (or $\bar{f}_j(t)$):
\begin{eqnarray}                 
     f_j(b) \;&=&\; T_j(\cos[\frac{\pi}{2}(\frac{b}{Bj})
                       ^{\mu_j}])^{\nu_j} \,,             \\
   \bar{f}_j(t)\;&=&\; B_j(\frac{2}{\pi}\arccos[(\frac{t}{T_j})
                                       ^{1/\nu_j}])^{1/\mu_j}  \,.
\end{eqnarray}

Each function will be determined by a triple of free parameters:
  \[ \delta_j,\varepsilon_j,d_j\,[fm]; \;\;\;h_j\,[fm^{-1}],\theta_j,c_j;                           \;\;\;\mu_j,\nu_j,T_j\,[GeV^2]. \]
There are then four free parameters more:
           \[ p_1, \;\;\; p_2, \;\;\; b_1\,[fm], \;\;\;b_2\,[fm], \]
 the values of which will be determined by fitting corresponding experimental data.

\section {Analysis of experimental data}

Many experiments concerning the measurement of elastic pp scattering at different energies have been performed. We will make use of the data obtained at $53\; GeV$ on ISR; see Ref. \cite{kun}. It will be shown that under the assumption of two different internal structures the full agreement may be obtained for the data of differential cross section in the interval $(-4.,0.)\, [GeV^2]$ of $t$. Some additional structures should be added when the data at higher values of $|t|$ should be interpreted.

\begin{table}[htbp]
	\centering
  \small
		\begin{tabular}{|c|ccc|}
			\hline
         $\delta_j$    & 0.920 & 1.45 & 0.444  \\
        $\varepsilon_j$ & 3.20  & 8.0  & 11.4  \\   
        $ d_j$         & 1.09 & 0.90  & 0.444  \\
 			\hline
         $h_j$         & 1.45  & 24.0   & 10.0   \\   
         $\theta_j$     & 6.81  & 1.53   & 1.34  \\
         $c_j$         & 0.20  & 1209.  & 8000.   \\
 			\hline
         $\mu_j$   & 3.28  & 0.940  & 1.07    \\
         $\nu_j$   & 0.148 & 0.150  & 0.041   \\
           $T_j$     & 236.7 &  23.0 & 68.8   \\
 			\hline
		\end{tabular}
	\caption{\it Values of free parameter triples determining  characteristics of different collision types}
	\label{tab:1}
\end{table}
                   
However, in pp elastic scattering the measured data are always obtained as a superposition of hadron and Coulomb scatterings. Thus, instead of Eq. (\ref{difs}) the equation
\begin{equation}
   \frac{d\sigma_{el}(t)}{dt}\,=\, \sum_1^3r_j\,F_j(t)\,.\, 10 \;\;+\;\; r_C\,F_C(t)                          \label{fct}    
\end{equation} 
must be used where (for the given energy) we can write
\begin{equation}
     F_C(t)\;=\; (\frac{53/137}{t})^2 \;.\; f_C(t)
\end{equation} 
and $f_C(t)$ is corresponding form factor that in the case of two colliding protons should be determined together with other proton characteristics. We shall write 
\begin{equation}    
  f_C(t) = (1\,-\,(1-e^{-(\xi_1|t|)^{\xi_2}})/(1+\xi_3 e^{-(\xi_1|t|)^{\xi_2}}) ; \end{equation}
where $\xi_j$ are free parameters. The parameter 
 $r_C$ in equation (\ref{fct}) represents the contribution of Coulomb scattering in addition to hadron scattering.      

The presented fit has been obtained with the help of the following values of the  basic free parameters:
  \[ p_1=0.565, \;\;\; p_2=0.309, \;\;\; r_C= 0.0017, \;\;\; b_1= 1.64\,[fm], 
                   \;\;\;b_2=1.42\,[fm]. \]
The form factor of Coulomb forces in the corresponding pp interaction is characterized by parameters
        \[  \xi_1= 0.958 , \;\;\; \xi_2= 8.50  , \;\;\; \xi_3= 11.28 . \]
The other parameters characterizing three individual proton collisions are then given in Tab. 1.

\begin{figure}[htb]
\begin{center}
\includegraphics*[scale=.42, angle=-90]{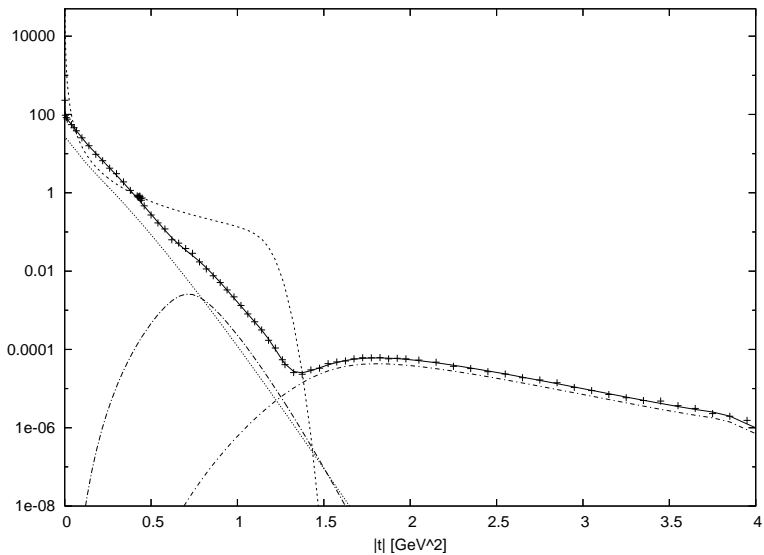}
\caption { \it {  Differential elastic cross section for pp scattering at 53 GeV; individual points - experimental data, full line - theoretical fit, dashed line - Coulomb interaction, other lines - contribution of individual collision types } }
 \end{center}
 \end{figure}

The result of the corresponding fit is shown in Fig. 1. Individual points represent the measured data of differential cross section (in milibarns) for elastic pp scattering at CMS energy of $53\, GeV$; the transverse momentum squared (in $GeV^2$) being shown on horizontal axis. The theoretical (fitted) dependence is represented by full line; good agreement being obtained in the whole considered interval of $t$. The dashed (upper) line represents the effect of Coulomb interaction (including the effect of corresponding form factors) while only a small fraction (given by $r_C$) contributes to measured values. The remaining lines represent the contributions of three individual collision types represented by functions $F_j(t)$ (determined in $fm^2$); it holds $F_j(0)=0$ in all three cases. 

\begin{figure}[t!]
\begin{center}
\includegraphics*[scale=.32, angle=-90]{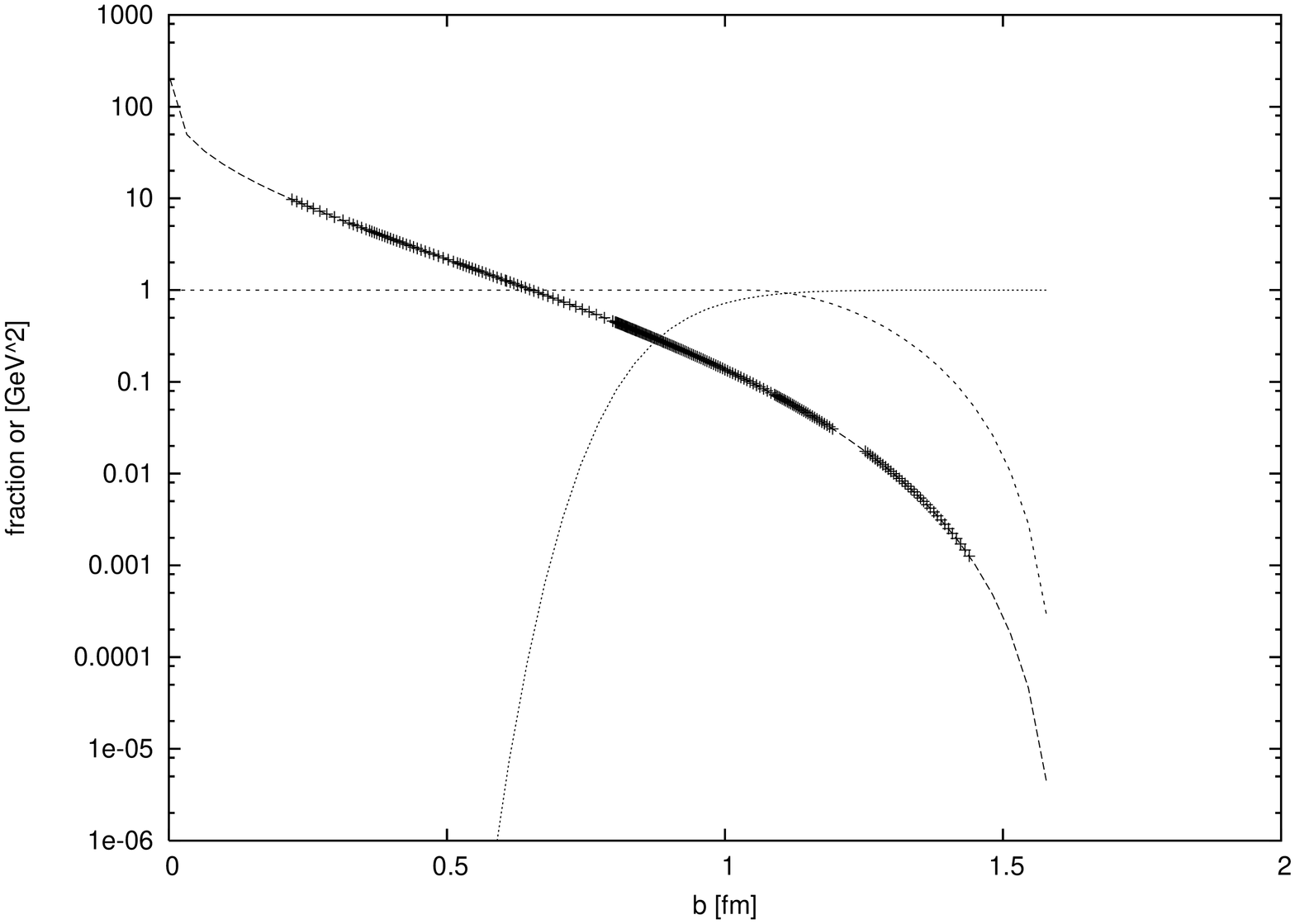}
\includegraphics*[scale=.32, angle=-90]{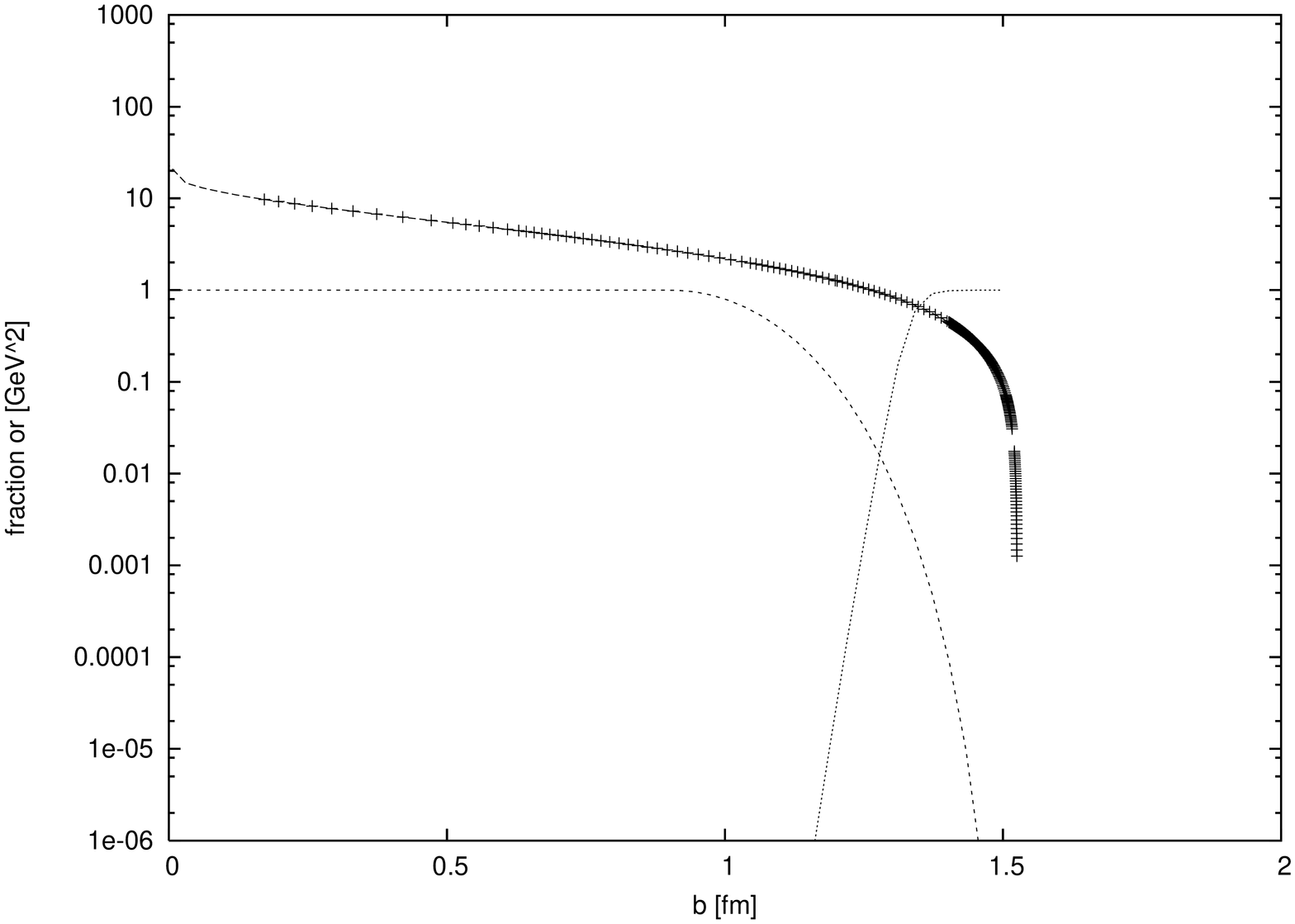}
\includegraphics*[scale=.32, angle=-90]{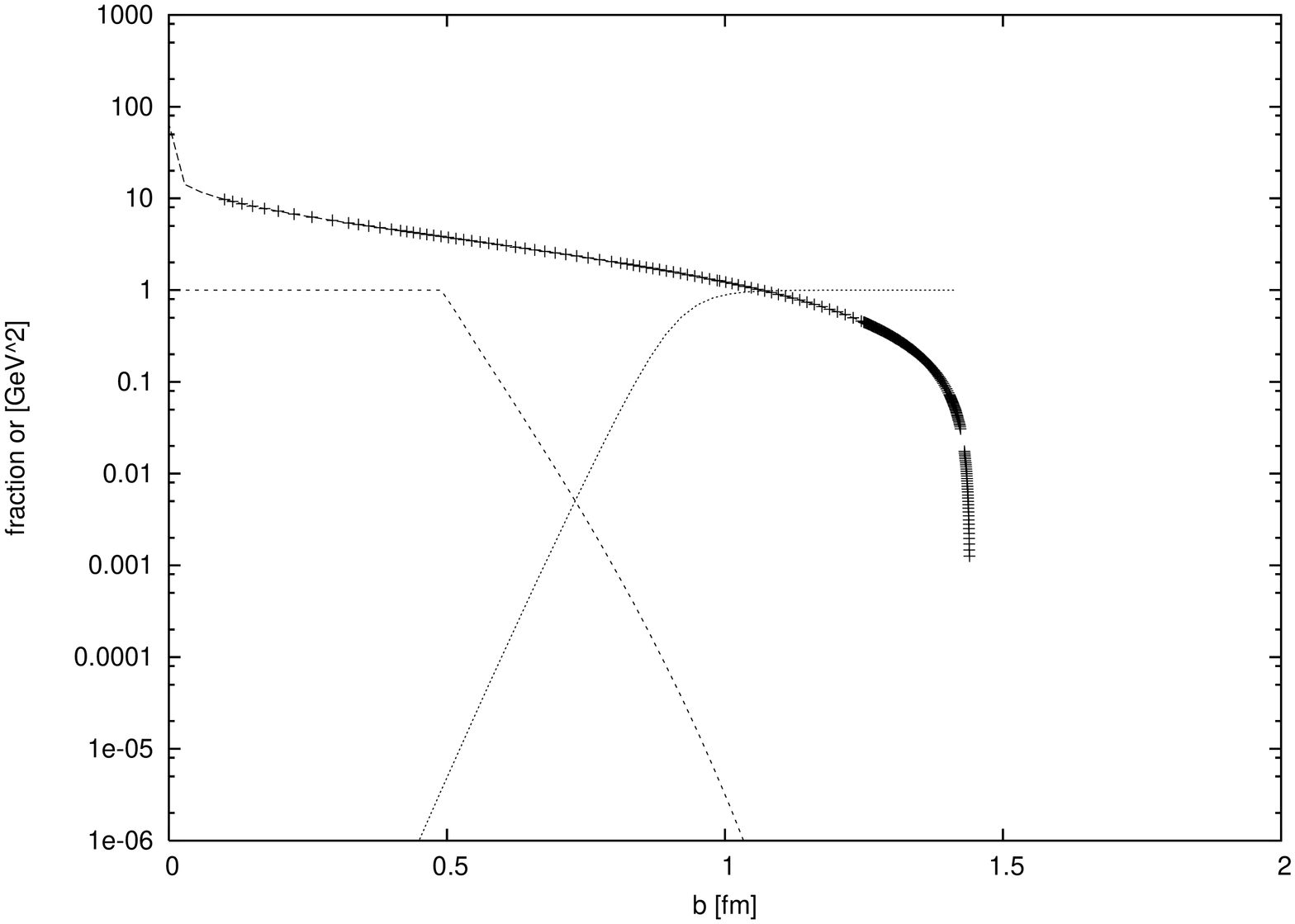}
  \caption { \it {Representation of functions $a_j(b)$ and $e_j(b)$ (dotted and dashed lines) and functions $f_j(b)$ together with experimental points in the correspondence to individual collision types } }
 \end{center}
 \end{figure}

As to the Coulomb form factor, it may be interesting to point to its values in the region of rather high values of $|t|$ (just under the dip) and then steep decrease. Such behavior may be explained by that the charged nucleons may execute the Coulomb scattering (and not hadronic collision) also at rather small distances due to ellipsoid form of individual structures. The greater Coulomb effect in the given region has been discovered already earlier in analyses based on eikonal models (significant interference contribution in the mentioned region); see Refs. \cite{kun,kun1}. 

The functions $a_j(b)$ and $e_j(b)$ (see Eqs. (\ref{ajb}) and (\ref{ejb})) representing the $b$-dependences of probabilities of total and elastic collisions are then shown in Fig. 2; see dotted and dashed lines.
The values of impact parameter are given on horizontal axis. The mean values (in $GeV^2$) of transverse momentum squared corresponding to individual values of $b$ are represented by individual points together with the curves representing $f_j(b)$ functions.

The total cross sections $\sigma_j$ for individual collision types may be then determined with the help of $a_j(b)$ functions; it holds
\begin{equation}
      \sigma_j\;=\; 2\pi\int_0^{B_j}\!db\,b\,a_j(b) \;.
\end{equation}
From the curves $a_j(b)$ obtained by fitting one gets then
     \[ \sigma_j\;=\;\;\; 54.2;\;\; 40.4; \;\; 7.46 \;\;mb, \]
which gives the average value $35.9\, mb$
for the hadron total cross section of two protons corresponding to two considered hadron structures (with given probabilities $p_1$ and $p_2$). The effect of other structures should be added if the whole measured differential cross section should be fitted. One must, of course, expect that the final value will remain always less than that being commonly considered (i.e., 43 mb - derived on the basis of optical theorem validity).     

As to the elastic cross section it may be obtained by integrating the measured differential cross section. One obtains in such a case
  \[ \sigma^{el}_{exp} = \int d\sigma_{el}(t)\;=\; 7.57\,mb . \]
The contributions of individual collision types may be obtained as
\begin{equation}
      \sigma^{el}_j\;=\; 2\pi\int_0^{B_j}db\,b\,a_j(b)\,e_j(b) \;.
\end{equation}
One obtains then numerically from fitted curves
 \[ \sigma^{el}_j\;=\;\;\; 22.97;\;\; 0.0074; \;\; 0.00045\;\;mb, \]
which gives average value $\sigma^{el}_{1-3} = 7.31\,mb$, being in good agreement with the experimentally determined value. 
\\

\section {Conclusion}

It has been shown that in the interval \mbox{$|t| \in (0.,4.)\, [GeV^2]$} the differential elastic nucleon cross section may be well fitted when one takes into account two different proton structures exhibiting the greatest dimensions (maximal values $1.64\; fm$ and $1.42\; fm$). Another structure with lower maximal dimension may be then responsible for collision processes corresponding to higher values of $|t|$, as it will be shown elsewhere.

The distribution of $|t|$ in individual collision types is characterized by the probability dependence increasing from zero if the impact parameter decreases from $b=b_{max}$.
It means that the optical theorem requiring the maximal probability  for $|t|=0$ cannot be applied to, which is in agreement with our recent results; see Ref. \cite{opt} where the validity of optical theorem in hadron collisions have been strongly impeached. 
It means that also all values of collision cross sections obtained in the past on the basis of such an assumption should be newly tested on more reliable grounds. 

  It is also the value $35.9\, mb$ obtained for total cross section that should be seriously analyzed and tested, being significantly lower than the value $43.\;mb$ commonly accepted. Our ontological probabilistic model might indicate that the value of ISR luminosity has been undervalued in corresponding ratio. And the same might hold for all colliders providing higher collision energies. 

{\footnotesize

\end{document}